\documentclass[iop,12pt,a4paper,showpacs,epsfigure,superscriptaddress]{iopart}
\usepackage{dcolumn}    
\usepackage{bm} 
\usepackage{graphicx}
\usepackage{latexsym}
\usepackage{amsfonts}   
\usepackage{amssymb}
\usepackage{array}      
\usepackage{epsfig}
\usepackage{color}

\begin{document}

\title{Transitionless quantum driving in open quantum systems}
\date{\today}

\author{G. Vacanti$^1$}
\address{$^1$Center for Quantum Technologies, National University of Singapore, 1 Science Drive 2, Singapore}

\author{R. Fazio$^{2,1}$}
\address{$^2$NEST, Scuola Normale Superiore \& Istituto di Nanoscienze-CNR, I-56126 Pisa, Italy}
\address{$^1$Center for Quantum Technologies, National University of Singapore, 1 Science Drive 2, Singapore}

\author{S. Montangero$^3$}
\address{$^3$Institut f\"ur Quanteninformationsverarbeitung, Universit\"at Ulm, 89069 Ulm, Germany}
\address{Center for Integrated Quantum Science and Technology (IQST), Universities of Ulm/Stuttgart and MPI for Solid State Research}

\author{G. M. Palma$^4$}
\address{$^4$NEST, Istituto Nanoscienze-CNR \& Dipartimento di Fisica e Chimica, Universita' degli Studi di Palermo,  
I-90123 Palermo, Italy}

\author{M. Paternostro$^5$}
\address{$^5$Centre for Theoretical Atomic, Molecular and Optical Physics, School of Mathematics and Physics, Queen's University, Belfast BT7 1NN, UK}
\author{V. Vedral$^{6,1,7}$}
\address{$^6$Clarendon Laboratory, University of Oxford, Parks Road,  Oxford 0X1 3PU, UK}
\address{$^1$Center for Quantum Technologies, National University of Singapore, 1 Science Drive 2, Singapore}
\address{$^7$Department of Physics, National University of Singapore, 3 Science Drive 4, Singapore}

\begin{abstract}
We extend the concept of superadiabatic dynamics, or transitionless quantum driving, to quantum open systems whose evolution is governed by a master equation in the Lindblad 
form. We provide the general framework needed to determine the control strategy required to achieve superadiabaticity. We apply our formalism to two examples consisting of a two-level system coupled to environments with time-dependent bath operators.

\end{abstract}

\maketitle
\date{\today}

The adiabatic theorem in quantum mechanics states that a physical system remains in the instantaneous eigenstate of the Hamiltonian that rules its dynamics, if a given perturbation is acting on it slowly enough~\cite{messiah,born1928}. The slower the time-dependence of the Hamiltonian the better the system is able to adapt to the corresponding changes. The implications of the adiabatic theorem have found key roles in the context of quantum computation~\cite{farhi2000}, in the physics of quantum phase transitions (see Ref.~\cite{dziarmaga2010} for a review), quantum ratchets, and pumping. 

Adiabatic dynamics is a way to control the evolution of the state of a quantum system through the time-dependence of  some Hamiltonian parameters, typically performed varying appropriately chosen external potentials. As perfect adiabaticity would require infinitely slow changes, the desired evolution can only be achieved approximately. In general, non-adiabatic corrections, although possibly very small, should thus be accounted for. 

At the opposite side of the spectrum lies optimal quantum control~\cite{walmsley2003}, which relies on the ability to engineer time-dependent Hamiltonians that allow to reach, in principle with unit fidelity, a given target state. Optimal quantum control~\cite{krotov1996,dalessandro2007,brif2010} has recently found very important applications in quantum information processing, where it has been shown to be crucial for the design of fast and high-fidelity  quantum gates~\cite{oqc_qip1,oqc_qip2,oqc_qip3,oqc_qip4,oqc_qip5}, the efficient manipulation of  simple quantum systems~\cite{machnes,machnes2,machnes3}, and the state preparation of quantum many-body systems~\cite{mb_qoc1,mb_qoc2}.

A very interesting connection between adiabatic dynamics and optimal control stems from a problem posed and  solved in Refs.~\cite{superadiab1,superadiab2,superadiab3,superadiab4}, and that can stated as follows: given a time-dependent Hamiltonian $\hat H_0(t)$  with instantaneous eigenstates $|\varphi_n(t)\rangle$, is it possible to identify  an additional term $\hat H_1(t)$ such that  the time dependent  Schr\"odinger equation driven by $\hat H(t) = \hat H_0(t)+\hat H_1(t)$ admits $|\varphi_n(t)\rangle$ as an  {\it exact} solution? With the provision of an explicit  construction of $\hat H_1(t)$  and the discussion of simple examples, Refs.~\cite{superadiab1,superadiab2,superadiab3,superadiab4} have  basically initiated a new field of investigations currently known as transitionless quantum driving, shortcut to adiabaticity or  superadiabatic dynamics. Protocols based on superadiabatic dynamics have been applied to a variety of different situations in atomic and molecular physics,  cold atomic systems, and many-body state engineering. The field has been recently reviewed in Ref.~\cite{torrontegui2012}, while the experimental  realisations have been reported for  artificial  two-level quantum system realised with Bose-Einstein condensates in optical lattices \cite{bason2012} and for nitrogen vacancies in diamonds ~\cite{zhang2013}.

To the best of our knowledge, the superadiabatic approach has only been considered for closed quantum systems (see however~\cite{ruschhaupt2012,Muga,Muga2}). Very recently, it was shown that when applied to quantum many-body systems, transitionless quantum driving may be achieved at the cost of highly non-local operations~\cite{DelCampo,DelCampo2}. Quite clearly, though,  a rigorous extension of the concept of superadiabaticity to open-system dynamics would be  much needed in order to enlarge the range of physical situations that can be addressed.


The provision of a framework for such generalisation is exactly the subject of this work. We reformulate the superadiabatic framework 
so as to adapt it to the case of an open-system dynamics written in a general Lindblad form. Our approach will be built on the definition  of open adiabatic dynamics as given in Ref.~\cite{Lidar} and will lead us to the statements given in Eqs.~(\ref{lindblad3}), 
(\ref{Lprime}) and (\ref{Ltqd}), which represent the main results of our work. We will then illustrate the effectiveness of our framework using  two examples involving the open dynamics of a single spin in a time-dependent environment.

\section{Unitary evolution}

In order to set the ground for the discussion on superadiabatic dynamics for open quantum system it is useful 
to  rephrase the results in Ref.~\cite{superadiab4} using a different  approach, which will be perfectly suited for a  generalization to the case of non-unitary evolutions.  

Let us consider a system spanning a Hilbert space of dimension $N$ and ruled by a time-dependent Hamiltonian $\hat H(t)$ with a discrete,  non-degenerate spectrum. By choosing the time-independent basis $\{|i\rangle\}$ ($i=\{1,\dots, N\}$), we can represent the Hamiltonian as  $\hat H(t)= \sum_{i,j} |i\rangle \langle i | \hat H(t) | j\rangle \langle j|$, and diagonalise it using the (time-dependent) similarity  transformation 
\begin{equation}
\hat U(t)=\sum_i{|\varphi_i (t) \rangle \langle i |},
\end{equation}
 where $|\varphi_i(t)\rangle$ is the $i^{\rm th}$ instantaneous eigenvector of $\hat H(t)$,  associated to the eigenvalue $E_i(t)$. It is straightforward to check that  $\hat U^{-1}(t) \hat H(t) \hat U(t)\equiv \hat H_{\rm d}(t) = \sum_i E_i(t) |i\rangle\langle i|.$

Following Ref.~\cite{Lidar}, let us now consider the time-dependent Schr\"odinger equation $\hat H(t)|\psi(t)\rangle = i \partial_t|{\psi}(t)\rangle$ and transform it to the picture defined by $\hat{U}^{-1}(t)$, which gives us 
\begin{equation}\label{schrodinger1}
[\hat H_{\rm d}(t) + i\partial_t{\hat U}^{-1}(t) \hat U(t)] |\psi \rangle_{\rm d} = i  \partial_t|{\psi}\rangle_{\rm d}
\end{equation}
with $|\psi \rangle_{\rm d} = \hat U^{-1} |\psi\rangle.$ By splitting $i\partial_t{\hat U}^{-1}(t) \hat U(t)$ into the sum of a diagonal term $\hat H_{\rm d}^\prime$ and an off-diagonal one $\hat H_{\rm nd}^\prime,$  Eq.~(\ref{schrodinger1}) is recast in the form
\begin{equation}
\label{schrodinger3}
	[\hat H_{\rm d}(t) + \hat H_{\rm d}^\prime(t) + \hat H_{\rm nd}^\prime(t)] |\psi \rangle_{\rm d} = i  \partial_t|{\psi}\rangle_{\rm d}.
\end{equation}
The explicit form of $\hat H_{\rm d}^\prime$ and $\hat H_{\rm nd}^\prime$ can be given as 
\begin{eqnarray}
	&\hat H_{\rm d}^\prime(t) = i\sum |i\rangle \langle i|\partial_t\hat{U}^{-1}(t) \hat U(t) |i\rangle \langle i|=
	i \sum  \langle \dot{\varphi}_i | \varphi_i \rangle |i\rangle \langle i|, \\
	&\hat H_{\rm nd}^\prime(t)= i\sum_{i\neq j} |i\rangle \langle i|\partial_t\hat{U}^{-1}(t) \hat U(t) |j\rangle \langle j| =
	 i \sum_{i\neq j} \langle \dot{\varphi}_i | \varphi_j \rangle | i \rangle \langle j|,
\end{eqnarray}
where we have dropped the explicit time dependence of the instantaneous eigenstates $\{|\varphi_i(t)\rangle\}$. While $\hat H_{\rm d}^\prime(t)$ encompasses  the contribution that leads to the geometric phases~\cite{shapere}, adiabaticity is enforced when $\hat H_{\rm nd}^\prime(t)$ is neglected. This can be easily seen by noticing  that both $\hat H_{\rm d}(t)$ and $\hat H_{\rm d}^\prime(t)$ are diagonal in the basis $\{| i \rangle\}$ and, by neglecting $\hat H_{\rm nd}^\prime(t)$, different eigenvectors will not be mixed across the evolution. 
In transitionless quantum driving, the goal is to find an additional term $\hat H_{\rm tqd}(t)$ such that the Schr\"odinger equation for the Hamiltonian $\hat H(t)+\hat H_{\rm tqd}(t)$  admits the adiabatic evolution of an eigenvector of $\hat H(t)$ as an exact solution. From the discussion above, it is straightforward to  see that such additional term is given by 
\begin{equation}
\label{Htqd}
	\hat H_{\rm tqd}(t) = -\hat U(t) \hat H^\prime_{\rm nd}(t) \hat U^{-1}(t).
\end{equation} 
Indeed, by applying $\hat U^{-1}$ to both side of the Schr\"odinger equation for the Hamiltonian $\hat H+\hat H_{\rm tqd}$, it is straightforward to get $[\hat H_{\rm d}(t) + \hat H_{\rm d}^\prime(t)] |\psi \rangle_{\rm d} = i  \partial_t|{\psi}\rangle_{\rm d}$. That is, the non-adiabatic term responsible for the coupling  between different eigenspaces of $\hat H(t)$, which is usually neglected in the adiabatic approximation, can be cancelled  exactly by adding the term $\hat H_{\rm tqd}(t)$ to the original Hamiltonian. Needless to say, the explicit calculation of $\hat H_{\rm tqd}(t)$  leads to the same expression given in Ref.~\cite{superadiab4}.

\section{Superadiabatic dynamics: Lindblad dynamics}

We are now in a position to generalize the framework discussed  above to the case of non-unitary evolutions. We will consider a general master equation in the Lindblad form $ \mathcal{L}[\varrho] = \dot{\varrho}$
for the density matrix $\varrho$ of the system. Here, $\mathcal{L}$ is the time-dependent  superoperator describing the non-unitary dynamics of the system and given by the general form
\begin{equation}
 	\mathcal{L}[\varrho] = -i [\hat H(t),\varrho] +\frac{1}{2}\sum_{j=1}^{N} (2 \hat\Gamma_j(t) \varrho \hat\Gamma_j^\dag(t) 
	- \{\hat\Gamma_j^{\dag}(t) \hat\Gamma_j(t),\varrho\})\label{LindbladOp}
\end{equation}
with $\hat H(t)$ the Hamiltonian of the system and $\hat\Gamma_j(t)$ the operators describing the system-environment interaction. Here $\{\cdot,\cdot\}$ stands for the anticommutator.  


The adiabatic dynamics in open system needs to be defined with care. In fact, due to the coupling of the system with the  environment, the energy-difference between neighbouring eigenvalues of the Hamiltonian no longer provides the natural time-scale  with respect to which a time-dependent Hamiltonian could be considered to be slow-varying. Here we follow the approach developed  in Ref.~\cite{Lidar}, according to which adiabaticity of open systems is reached when the evolution of the state of a system occurs without mixing  the various Jordan blocks into which ${\cal L}$ can be decomposed.  {The use of Jordan block decomposition is necessary due to the fact that the Lindblad operator ${\cal L}$ might not be diagonalizable  in general. Although many important problems deal with diagonalisable Lindblad superoperators, a general treatment of transitionless quantum driving in open systems requires the Jordan formalism. Explicit {\it ad hoc} examples of non-diagonalizable Lindblad superoperators can be constructed even for simple systems such a single qubit, as shown in~\cite{Lidar}. Although for the sake of our analysis it is the general formalism to be relevant, we stress that the search for less contrived instances is the topic of current studies.}

Equipped with this definition we are now ready to describe superadiabatic dynamics of open systems. In order to use the formalism introduced above for the case of pure states undergoing a unitary evolution, we need to write all superoperators as matrices  and all density matrices as vectors. Following Ref.~\cite{Lidar,alickilendi}, we start  by defining a time independent basis in the $D^2$-dimensional space (where $D$ is the dimension of the Hilbert space) of the density matrices as $\hat B\equiv \{\hat\sigma_i\}$. This could consist, for example, the three Pauli matrices  and the identity matrix in the case of a single spin-$1/2$. Once we have defined the basis $\hat B,$ the density  matrix can be transform into a ``coherence vector'' living in a $D^2$-dimensional space as  $|\varrho\rangle \rangle =(\rho_1, \rho_2, \dots, \rho_{D^2})^\dag,$ where $\rho_j= \rm{Tr}[\hat\sigma_j^\dag \varrho].$ On the other hand, the Lindblad superoperator $\mathcal{L}$ becomes a $D^2\times D^2$ time-dependent matrix $L(t)$ (which we will call a ``supermatrix") whose elements are given by  $L_{jk}(t)= \rm{Tr}[\hat\sigma_j^\dag (\mathcal{L}_t[\hat\sigma_k])].$ With this notation, the master equation now reads 
\begin{equation}
\label{LindbladEq2}
 	L(t) |\varrho \rangle\rangle = |\dot{\varrho}\rangle\rangle.
\end{equation}
Although  the supermatrix $L(t)$ might be  non-Hermitian, in which case it cannot be diagonalized in general, it is  always possible to find a similarity transformation $C(t)$ such that  $L(t)$ is written in the canonical Jordan form 
\begin{equation}
	L_{\rm J}(t)=C^{-1}(t) L(t) C(t)= \rm{diag}[J_1(t),\dots,J_N(t)],
\label{JordanDec}
\end{equation}
where  $J_{\nu}(t)$ represents the Jordan block (of dimension $M_\nu$) corresponding to the  the eigenvalue $\lambda_{\nu}(t)$ of $L(t).$  The number $N$ of Jordan blocks is equal to the number of linear independent eigenvectors of $L(t)$ and the similarity transformation is given by  
\begin{equation}
\label{C}
	C(t)= \sum_{\nu=1}^{N}\sum_{\mu_{\nu}=1}^{M_\nu} |\mathcal{D}_{\nu,\mu_{\nu}}(t)
	\rangle\rangle \langle \langle \sigma_{\nu,\mu_\nu}|,
\end{equation}
where $\{|\mathcal{D}_{\nu,\mu_{\nu}}(t)\rangle\rangle\}$ is a basis of right instantaneous quasi-eigenvectors of $L(t)$ associated with the eigenvalues $\{\lambda_\nu(t)\}.$ {The set of right quasi-eigenstates $\{ |\mathcal{D}_{\nu,\mu_{\nu}}(t)\rangle\rangle\}$ is defined through the equation
\begin{equation}\label{quasieigevectorsEq}
	L(t) |\mathcal{D}_{\nu,\mu_{\nu}}(t)\rangle\rangle= |\mathcal{D}_{\nu,\mu_{\nu}-1}(t)\rangle\rangle 
	+ \lambda_{\nu}(t) |\mathcal{D}_{\nu,\mu_{\nu}}(t)\rangle\rangle,
\end{equation}
where $|\mathcal{D}_{\nu,0}(t)\rangle\rangle$ represents  the eigenvector of $L(t)$ corresponding to the eigenvalue 
$\lambda_\nu(t)$ and $\mu_{\nu} = \{1,\dots,M_{\mu}\},$ with $M_{\mu}$ the dimension of block $J_{\mu}.$}
On the other hand,  $\{\langle \langle \sigma_{\nu,\mu_{\nu}} |\}$  are the vector of the basis $B$ introduced above with the index $i$ now defined as $i=\sum_{k=0}^{\nu-1}M_k+\mu_{\nu}$ ($M_0=0$). The inverse transformation $C^{-1}(t)$ (such that $C^{-1}(t)C(t)=C(t)C^{-1}(t)=\mathcal{I}$) can be defined in a conceptually analogous way by considering the set of left instantaneous quasi-eigenvectors of $L(t)$. As the set $\{|\mathcal{D}_{\nu,\mu_{\nu}}(t)\rangle\rangle\}$ embodies the basis where $L(t)$ is in Jordan form, we immediately get  that $L_J(t) = C^{-1}(t)L(t)C(t)$. Needless to say, when $L(t)$ is diagonalizable the same arguments and definitions above apply with $M_{\nu}$ becoming the multiplicity of the eigenvalue  $\lambda_{\nu}$ and right (left) quasi-eigenvectors being promoted to the role of exact right (left) eigenvectors of $L(t).$

Exploiting the formal equivalence between Eq.~(\ref{LindbladEq2}) and the (imaginary-time) Schr\"odinger equation for non-Hermitian Hamiltonians, the same  arguments illustrated above in the context of unitary evolutions can be used here. We thus apply the  transformation $C^{-1}(t)$  to both side of Eq.~(\ref{LindbladEq2}). After some straightforward manipulation, the latter is rewritten in the form  
\begin{equation}
\label{lindblad3}
	[L_{\rm J}(t) + L_{\rm J}^\prime(t) + L_{\rm nd}^\prime(t)] |\varrho \rangle \rangle_{\rm J} 
	=  |\dot{\varrho}\rangle\rangle_{\rm J}
\end{equation}
which is analogous to Eq.~(\ref{schrodinger3}) and where we have introduced 
\begin{eqnarray}
\label{Lprime}
	&L_{\rm J}^\prime(t)= \sum |\sigma_{\nu,\mu_{\nu}} \rangle \rangle{\cal C}^{\nu,\nu}_{\mu_\nu,\mu_{\nu}}(t) \langle \langle \sigma_{\nu,\mu_{\nu}}|, \\
	& L_{\rm nd}^\prime(t)= \sum_{\nu\neq \nu^{\prime}}  |\sigma_{\nu,\mu_{\nu}}\rangle\rangle  
	{\cal C}^{\nu,\nu'}_{\mu_\nu,\mu'_{\nu'}}(t) \langle \langle \sigma_{\nu^\prime,\mu_{\nu^\prime}^\prime}| 
\end{eqnarray}
with ${\cal C}^{\nu,\nu'}_{\mu_\nu,\mu'_{\nu'}}(t)=\langle\langle \sigma_{\nu,\mu_{\nu}}|\dot{C}^{-1}(t) C(t) |\sigma_{\nu^\prime,\mu_{\nu^\prime}^\prime} \rangle\rangle$. In both Eq.~(\ref{lindblad3}) and (\ref{Lprime}), the pedex ${\rm J}$ indicates that the matrix $L(t)$ is in the Jordan form and  the coherence  vectors are transformed as $|\varrho \rangle \rangle_{\rm J} = C^{-1}|\varrho \rangle \rangle$. 

For open systems, the problem of transitionless quantum driving consists of finding an additional term $L_{\rm tqd}(t)$ such that different  Jordan blocks of $L$ alone evolve independently under the action of $L(t)+L_{\rm tqd}(t)$. Since the two terms  $L_{\rm J}(t)$ and $L_{\rm J}^\prime(t)$ preserve the Jordan blocks structure, any  admixture between different Jordan  blocks is bound to arise from $L_{\rm nd}^\prime(t)$. Therefore, by using the same approach sketched in the unitary case, we can infer the form of the additional term $L_{\rm tqd}(t)$ as 
\begin{equation}
\label{Ltqd}
	L_{\rm tqd}(t)= -C(t)L_{\rm nd}^\prime(t) C^{-1}(t).
\end{equation}
Eq.~(\ref{Ltqd}) extends and generalizes the result valid for the unitary case [cf. Eq.~(\ref{Htqd})] to quantum open-system dynamics and is the main result of this work. Just like in the unitary case, $L_{\rm tqd}(t)$ encompasses the control that should be implemented so that the state of the system remains, across the evolution, in an instantaneous eigenstate. The required control term could be either on the unitary part of the dynamics (i.e. an additional Hamiltonian term), or in the non-unitary one, which would require the engineering of a proper quantum channel. While  we identify a physically relevant condition that ensures that the correction term is of Hermitian nature in the following paragraph, in the latter case there is no guarantee that the correction adds up to the dynamics of the system so as to give a completely positive map~\cite{exem}. When this is the case, though, it is sufficient to add an {\it effective damping} term diagonal in the correction term, large enough to re-instate complete positivity. 

{It is worth noting that, analogously to the case of adiabatic unitary dynamics,  the term $L^\prime_{nd}$ cancels exactly the terms in the evolution that would be neglected when the adiabatic approximation is enforced. By differentiating Eq.~(\ref{quasieigevectorsEq}) it is possible to explicitly link the correction term $L^\prime_{nd}$ to the neglected terms under the adiabatic approximation. For example, for unidimensional Jordan blocks (i.e. for a fully diagonalisable Lindblad operator with non-degenerate spectrum) we can write the off-diagonal  matrix elements of the correction term as~\cite{Lidar}
\begin{equation}
\langle\langle \dot{\mathcal{D}}_{i}(t)| \mathcal{D}_{j}(t) \rangle\rangle = \frac{\langle \langle \mathcal{D}_{i}(t) | \dot{L}(t) | \mathcal{D}_{j}(t) \rangle \rangle}{\lambda_j-\lambda_i}.
\end{equation}
The general case of non-trivial Jordan block can be treated analogously, although the correction term would assume a more complicated (although conceptually equivalent) expression (cf. Ref.~\cite{Lidar} for more details about the adiabatic approximation in open systems).}

We now address the question of whether is possible to provide a necessary condition for the Hermitian nature of the correction term in Eq.~(\ref{Ltqd}) is always Hermitian. Let us now consider a Lindblad superoperator on the form
\begin{equation}\label{LindbladUGen}
\mathcal{L}[\rho] =  \sum_k \frac{\gamma_k}{2}[2 \hat \Gamma_k(t) \rho \hat \Gamma^\dag_k(t)  - \{\hat \Gamma^\dag_k(t) \hat \Gamma_k(t),\rho \}],
\end{equation}
where we assumed $\hat \Gamma_k(t) =  \hat U^\dag(t) \hat \Gamma_0^{k}  \hat U(t)$ for a given global unitary operator $\hat U(t)$ and  time-independent jump operators $\hat \Gamma_0^{k}.$  By moving to a rotating frame defined by $\hat U(t)$ and calling $\tilde\rho= \hat U(t)\rho(t) \hat U^\dag(t)$ the density matrix in such a frame, we get the Lindblad equation 
\begin{equation}
\dot{\tilde{\rho}} = \sum_k \frac{\gamma}{2}[2  \hat \Gamma_0^{k} \tilde{\rho} \hat \Gamma_0^{k}  - \{\hat \Gamma_0^{k} \hat \Gamma_0^{k},\tilde{\rho} \}] - i[i \dot{\hat U}(t) \hat U^\dag(t),\tilde{\varrho}].
\end{equation}
That is, in the rotating frame generated by $\hat U(t),$ the time dependence of the Lindblad operator is cancelled, and different eigenvectors will evolve independently. This simple argument shows that, whenever the non-unitary part of the evolution of a system is governed by jump operators such as $\Gamma_k(t)$, the superadiabatic correction is provided by the Hamiltonian term $ \hat H_{\rm tqd}(t) = i\dot{\hat U}(t) \hat U^\dag(t).$  A more formal proof is given in Appendix~\ref{supplem}.

\section{Examples}\label{secexamples}

In order to illustrate the general formalism described above, let us now discuss some simple examples involving a single-spin system. The first addresses the case of a single spin affected by a dissipative mechanism described by the super operator
\begin{equation}
\label{adlindblad}
\mathcal{L}_{\rm ad} [\varrho] =  \frac{\gamma}{2}[2 \hat\sigma^-_{\bm n} \varrho \hat\sigma^+_{\bm n}  - \{\hat\sigma^+_{\bm n} \hat\sigma^-_{\bm n},\varrho \}]
\end{equation}
$\hat\sigma^-_{\bm n} =(\hat\sigma^+_{\bm n})^\dag= |{\downarrow}\rangle_{\bm n} \langle{\uparrow}|$ the lowering ladder operator along the direction ${\bm n}$, and $\{|{\downarrow}\rangle,|{\uparrow}\rangle\}$ the two spin states of the system. The dissipation occurs along a direction in the single-spin Bloch sphere identified by the unit vector ${\bm n}=(\sin{\theta} \cos{\phi}, \sin{\theta} \sin{\phi}, \cos{\theta})$ with $\theta$ and $\phi$ the azimuthal and equatorial angle, respectively. In order to write explicitly both the Liouvillian supermatrix $L(\theta,\phi)$ and the corresponding coherence vector, we choose the ordered basis $\hat B \equiv (\hat\mathcal{I},\hat\sigma_x,\hat\sigma_y,\hat\sigma_z)$.  Let us now consider the case in which the direction of the dissipation ${\bm n}$ precesses around the $z$ axis of the Bloch sphere at a constant angular velocity $\omega$, maintaining  a fixed azimuthal angle $\theta_0$, and a constant damping rate $\gamma$. By setting $\phi=\omega t$ and employing the result in Eq. (\ref{Ltqd}), we can find the explicit form of the $4\times4$ supermatrix $L_{\rm tqd} (\theta,\omega t)$ required to achieve superadiabaticity in this example. An explicit calculation shows that a purely Hamiltonian contribution of the form $\mathcal{L}_{\rm tqd}[\varrho] = -i[\hat H_{\rm tqd}(t),\varrho]$ with $\hat H_{\rm tqd}(t) = ({\bm n}\times \dot{\bm n})\cdot \hat{\bm\sigma}$, is sufficient to achieve superadiabaticity. Indeed, the correction term is a magnetic field  which at any instant  induces a rotation that cancels the time-dependence of the original Lindblad superoperator. Being Eq.~(\ref{adlindblad}) a particular case of the more general expression in Eq.~(\ref{LindbladUGen}), the correction term corresponds to $\hat H_{\rm tqd}(t) = i\dot{\hat U} \hat U^\dag,$ as expected.

 Let us now consider a simple example involving two qubits. We start by  designing a Lindblad operator which generate a time evolution map whose fix point is a Bell state $|\psi\rangle = (1/\sqrt{2})(|00\rangle  + |11\rangle).$ Such state can be obtain by applying a unitary operation $\hat U$ to the state $|00\rangle,$ where $\hat U$ represent an Hadamard transformation on one of the qubit followed by a C-NOT gate. The operation $\hat U$ can be represented by the $4\times4$ matrix 
\begin{equation}\label{entanglingU}
\hat U = \frac{1}{\sqrt{2}} 
\left(
\begin{array}{cccc} 
1 & 0& 1 & 0\\ 
0& 1 & 0 &1\\
0& 1 & 0& -1\\
1& 0 & -1&0\end{array}\right)
\end{equation}
The Lindblad map having the state $|\psi\rangle$ as a fix point  has the form given in Eq.~(\ref{LindbladUGen}) with jump operators 
\begin{equation}
\Gamma_1 = \hat U \big(|0\rangle_1 \langle 1| \otimes \hat\mathcal{I}_2\big) \hat U^{\dag}; \quad \Gamma_2 = \hat U \big(\hat\mathcal{I}_1 \otimes |0\rangle_2 \langle 1| \big) \hat U^{\dag}
\end{equation}

Let us now consider a  unitary operation
\begin{equation}\label{U(t)}
\hat U(t) = \left(\begin{array}{cccc} \cos{\theta} & 0& e^{-i\phi} \sin{\theta} & 0\\ 
0& \cos{\theta} & 0 &e^{-i\phi} \sin{\theta}\\
0& e^{i \phi} \sin{\theta} & 0& -\cos{\theta}\\
e^{i \phi} \sin{\theta}& 0 & -\cos{\theta}&0\end{array}\right)
\end{equation}
This unitary operation represents a generalization of the one given in Eq. (\ref{entanglingU}) in which the Hadamar transformation is substituted by a general rotation specified by the angles $\phi$ and $\theta.$ The case we are interested in is the one in which such angles are time-dependent. For simplicity, we assume $\phi=0,$ so the only time-dependent parameter is $\theta(t).$ This means that the Jump operators $\Gamma_k(t)$ are now time dependent, with the time dependence included in the parameter $\theta(t).$

The scenario we consider is the following: we consider a Lindblad whose fix point is a particular state, for example $|\psi(t_0)\rangle = (1/ \sqrt{2})(|00\rangle + |11\rangle),$ which correspond to $\theta(t_0) = \pi/4,$ with $t_0$ the time at which the system has reached such state. At this point, we can change the parameter $\theta,$ and consequently the jump operators $\Gamma_k(t).$ In such a way, the stationary state of the system can be dragged from the initial state 
$|\psi(t_0)\rangle = (1/ \sqrt{2})(|00\rangle + |11\rangle)$ to the state $|\psi(t)\rangle = (\cos{\theta(t)}|00\rangle + \sin{\theta(t)}|11\rangle)$ at time $t.$ If the changes in the parameter $\theta(t)$ are slow, the system will remain in the instantaneous fix point at all times $t$ with good approximation. On the other hand, by implementing the super-adiabatic protocol for open systems, we can change the prepared state  exactly and without the constrain of slowly changing jump operators. 

In this particular example, the super-adiabatic  correction  needed to obtain an exact driving can be easily calculated as $i \dot{\hat U}(t) \hat U^\dag(t).$ Using Eq. (\ref{U(t)}) with $\phi=0,$ the correction is given by 
\begin{equation}\label{UdotUdag}
H_{\rm tqd} = i\dot{\hat U}\hat U^{\dag} = \left(\begin{array}{cccc}0 & 0& 0& -i\dot{\theta}\\ 
0& 0 & -i\dot{\theta} &0\\
0&i\dot{\theta} & 0&0\\
i\dot{\theta}& 0 & 0&0\end{array}\right)
\end{equation}
which can be written as 
\begin{equation}\label{Htqd}
H_{\rm tqd} = -i\dot{\theta} (|00\rangle\langle 11| + |01\rangle\langle 10|) + h.c.
\end{equation}

\section{Conclusions}

We have proposed the extension of  superadiabatic dynamics to systems undergoing an explicitly open evolution. Although we have considered, for the sake of simplicity, examples involving only a small number of spins, the method that we have proposed is entirely general and can indeed be applied to instances of more complex systems. For example, we foresee that superadiabatic techniques for open system will play a key role in the context of dissipative quantum state engineering~\cite{DQSE,DQSE2,DQSE3,DQSE4,DQSE5,Buchleitner2013} and in the emerging field of thermodynamics of quantum systems. A promising result in this sense is provided by Ref.~\cite{John}, where the design of superadiabatic quantum engines has been reported.  Moreover, {in general,  the class of problems for which the time-dependent  Lindblad superoperator admits one non-degenerate Jordan block with eigenvalue $\lambda_0=0$ for any $t$ is of particular interest in the context of transitionless quantum driving. Indeed, in this cases the system admit a unique stationary state for any time. The correction term, in such case, can be seen as the one needed to keep the system in its exact stationary state throughout the whole evolution.}

\section*{Acknowledgments}

We would like to thank  V. Giovannetti, J. Goold, and A. Monras for useful discussions. We acknowledge the Ministry of Education of Singapore, the DFG via SFB/TRR21, the EU (IP-SIQS, TherMiQ), the PRIN-MIUR, the UK EPSRC (EP/G004579/1) and the John Templeton Foundation (grant ID 43467) for financial support. VV is a fellow of Wolfson College Oxford and is supported by the John Templeton Foundation and the Leverhulme Trust (UK).

\section*{Appendix}
\label{supplem}

For simplicity, we assume that the Lindblad operator is diagonalizable. This is the case considered by Kraus {\it et. al.} in the context of  quantum state preparation of a chain of qubits~\cite{DQSE3}. However,  the proof can be generalize to the case in which the Lindblad admits only a Jordan block decomposition.

Let us consider a Lindblad operator in the general form given in Eq.~(\ref{LindbladUGen})
with the jump operators given by $\Gamma_k(t) = \hat U^{\dag}(t) \Gamma_k^{0} \hat U (t).$ Following Ref.~\cite{DQSE3}, we notice that the problem of finding the instantaneous eigenstates of Eq.~(\ref{LindbladUGen}) can be reduced to the problem of finding the eigenstates for the time independent Lindblad operator at time $t_0.$ Let us denote  by  $\{|\mathcal{D}^{(0)}_n\rangle\rangle\}$ the eigenstates of the time independent Lindblad operator $\mathcal{L}_{\{\Gamma_k^{0}\}}$ and by  $\{D^{(0)}_n\}$ the corresponding eigen-matrices, i.e.
\begin{equation}
\mathcal{L}_{\{\Gamma_k^{0}\}}[D^{(0)}_n] =  \lambda_n D^{(0)}_n.
\end{equation}
 The set of matrices $\{D^{(0)}_n\}$ forms a basis in  $\mathcal{B}(\mathcal{H})$ (this is also true for the set of quasi-eigenstates of $\mathcal{L}_{\{\Gamma_k^{0}\}}$ when the Lindblad operator is not diagonalizable). Moreover, the  eigenstates of the time-dependent Lindblad operator $\mathcal{L}_{\{\hat U^{\dag}(t) \Gamma_k^{0} \hat U (t)\}},$ which we will denote by  $\{|\mathcal{D}_n(t)\rangle\rangle\}$ and  $\{D_n(t)\},$ can be found as~\cite{DQSE3}
\begin{equation}\label{EigenstatesL}
D_n(t) = \hat U^{\dag}(t) D^{(0)}_n \hat U (t).
\end{equation}
As above, this relation also hold in the case of non-diagonalizable Lindblad operators for the quasi-eigenstates. Eq.~(\ref{EigenstatesL}) gives the important link between the eigenvectors of $\mathcal{L}$ at the initial time $t_0$ (indeed, we  choose the initial time such that $\hat U(t_0) = \mathcal{I}$) and the eigenstates at a generic time $t.$ 

Let us now prove that the general correction term $\dot{C}^{-1}C$ corresponds to an  Hamiltonian term  $i \dot{\hat U}\hat U^{\dag}$  if the jump operators can be written in the form $\Gamma_k(t) = \hat U^{\dag}(t) \Gamma_k^{0} \hat U (t).$ Since we are allowed to choose any time-independent basis for describing the system in Banach space, we pick the basis of quasi-eigenvectors of $\mathcal{L}$ at time $t_0,$ i.e. the set of matrices $\{D^{(0)}_n\}.$ We then have that 
\begin{eqnarray}
\label{C-1dotC}
&C=\sum_{n} |\mathcal{D}_n(t)\rangle\rangle\langle \langle \mathcal{D}_n^{(0)}|, \quad \quad C^{-1}=\sum_{n} |\mathcal{D}_n^{(0)}\rangle\rangle\langle \langle \mathcal{D}_n(t) |,\\ 
&\dot{C}^{-1} C = \sum_{n,m} \langle \langle \dot{\mathcal{D}}_n(t) |\mathcal{D}_m(t)\rangle\rangle  |\mathcal{D}_n^{(0)}\rangle\rangle\langle \langle \mathcal{D}_m^{(0)}|.
\end{eqnarray}
By definition, the matrix $\dot{C}^{-1} C$ corresponds to a superoperator  $\mathcal{C}(\rho)$ through the relation 
\begin{equation}\label{C-1dotCdef}
[\dot{C}^{-1} C]_{i,j} = \rm{Tr}\{D_i^{(0) \dag} \mathcal{C}(D_j^{(0)})\}.
\end{equation}
From Eq. (\ref{C-1dotC}), we also have 
\begin{equation}
[\dot{C}^{-1} C]_{i,j} = \langle \langle \dot{\mathcal{D}}_i(t) |\mathcal{D}_j(t)\rangle\rangle.
\end{equation}
This can be written in terms of density matrices as 
\begin{equation}
[\dot{C}^{-1} C]_{i,j} = \rm{Tr}\{\dot{D}_i^\dag(t) D_j(t) \}.
\end{equation}
Using Eq. (\ref{EigenstatesL}), this can be written as 
\begin{eqnarray}
\label{c-1dotccalc}
[\dot{C}^{-1} C]_{i,j} &=\rm{Tr}\{(\dot{\hat U}^\dag D_i^{(0)\dag}\hat U + \hat U^\dag D_i^{(0)\dag} \dot{\hat U}) \hat U^\dag D_j^{(0)} \hat U \}=\\ 
&=  \rm{Tr}\{(\dot{\hat U}^\dag D_i^{(0)\dag}\hat U \hat U^\dag D_j^{(0)} \hat U\} + \rm{Tr}\{ \hat U^\dag D_i^{(0)\dag} \dot{\hat U} \hat U^\dag D_j^{(0)} \hat U \}=\\ 
&= \rm{Tr}\{(D_i^{(0)\dag} D_j^{(0)} \hat U \dot{\hat U}^{\dag}\} + \rm{Tr}\{ D_i^{(0)\dag} \dot{\hat U} \hat U^\dag D_j^{(0)} \}=\\
&= \rm{Tr}\{(D_i^{(0)\dag} (D_j^{(0)} \hat U \dot{\hat U}^{\dag} + \dot{\hat U} \hat U^\dag D_j^{(0)}) \}.
\end{eqnarray}
As $\hat U$ is unitary, we have that 
\begin{equation}
0 = \dot{\mathcal{I}} = \dot{(\hat U \hat U^\dag)} = \dot{\hat U}\hat U^{\dag} + \hat U \dot{\hat U^\dag} \quad \Longrightarrow \quad  \hat U \dot{\hat U^\dag} = - \dot{\hat U}\hat U^{\dag}.
\end{equation}
Substituting in Eq.~(\ref{c-1dotccalc}), we obtain 
\begin{equation}
[\dot{C}^{-1} C]_{i,j} = \rm{Tr}\{(D_i^{(0)\dag} (\dot{\hat U} \hat U^\dag D_j^{(0)} - D_j^{(0)} \dot{\hat U}\hat U^{\dag} ) \}.
\end{equation}
Using the definition given in Eq. (\ref{C-1dotCdef}), the superoperator $\mathcal{C}(\rho)$ corresponding to $\dot{C}^{-1}C$ is then given by 
\begin{equation}
\mathcal{C}(\rho) = -i [i\dot{\hat U}\hat U^\dag, \rho].
\end{equation}


\begin{thebibliography} {99}
\bibitem{exem} This is the case, for instance, for a spin-$1/2$ particle subjected to a magnetic filed along the $z$ axis and a dephasing mechanism rotating at frequency $\omega$ in the $z=0$ plane. In such instance, it is straightforward to see that the superadiabatic correction is not Hermitian and that the Kossakowski matrix of the overall master equation is, in general, not positive definite.
\end{thebibliography}
\end{document}